\documentclass[pra,twocolumn]{revtex4-1}
\usepackage{float,graphicx,color,amsmath,
amscd,amsfonts,amssymb}
\begin{document}
\title{Quantum key distribution
protocol based on contextuality monogamy}
\author{Jaskaran Singh}
\email{jaskarasinghnirankari@iisermohali.ac.in}
\affiliation{Department of Physical Sciences,
Indian
Institute of Science Education \&
Research (IISER) Mohali, Sector 81 SAS Nagar,
Manauli PO 140306 Punjab India.}
\author{Kishor Bharti}
\email{ms11016@iisermohali.ac.in}
\affiliation{Department of Physical Sciences,
Indian
Institute of Science Education \&
Research (IISER) Mohali, Sector 81 SAS Nagar,
Manauli PO 140306 Punjab India.}
\author{Arvind}
\email{arvind@iisermohali.ac.in}
\affiliation{Department of Physical Sciences,
Indian
Institute of Science Education \&
Research (IISER) Mohali, Sector 81 SAS Nagar,
Manauli PO 140306 Punjab India.}
\begin{abstract}
The security of quantum key distribution(QKD)
protocols hinges upon features of physical systems
that are uniquely quantum in nature. We explore
the role of quantumness as qualified by quantum
contextuality, in a QKD scheme.  A
new QKD protocol based on the
Klyachko-Can-Binicio\ifmmode \breve{g}\else \u{g}\fi{}lu-Shumovsky(KCBS) contextuality
scenario using a three-level quantum system is
presented. We explicitly show the unconditional
security of the protocol by a generalized
contextuality monogamy relationship based on
the no-disturbance principle. This protocol
provides a new framework for QKD  which has
conceptual and practical advantages over other
protocols.  
\end{abstract}
\maketitle
\section{Introduction}
\label{intro}
The existence of pre-defined values for quantum
observables that are independent of any
measurement settings, has been a matter of debate
ever since quantum theory came into existence.
While Einstein made a case for looking for hidden
variable theories that would give such
values~\cite{EPR}, the work of John Bell proved
that such local hidden variable theories cannot be
compatible with quantum
mechanics~\cite{Bell_theorem}. This points
towards a fundamental departure of the behaviour of
quantum correlations from the ones that can be
accommodated within classical descriptions. While
the departure from classical behaviour indicated by
Bell's inequalities requires composite quantum
systems and the assumption of locality, the
contradiction between assignment of predefined
measurement-independent values to observables and
quantum mechanics, goes deeper and was brought out
more vividly by the discovery of quantum
contextuality~\cite{Kochen1967}. In a
non-contextual classical description, a joint
probability distribution exists for the results of
any joint measurements on the system, and the
results of a measurement of a variable do not
depend on other compatible variables being
measured. Quantum mechanics precludes such a
description of physical reality; on the contrary
in the quantum description, there exists a context
among the measurement outcomes, which forbids us
from arriving at joint probability distributions
of more than two observables. Given a situation
where an observable $A$ commutes with two other
observables $B$ and $C$ which do not commute with
each other: a measurement of $A$ along with $B$
and a measurement of $A$ along with $C$, may lead
to different measurement outcomes for $A$. Thus,
to be able to make quantum mechanical predictions
about the outcome of a measurement, the context of
the measurement needs to be specified.

The first proof that the quantum world is
contextual, 
was given by Kochen
and Specker
and involved 117 different vectors in a
3-dimensional Hilbert space~\cite{Kochen1967}. Subsequently, the
number of observables required for such a
`no-coloring' proof was brought down to 31 by 
Conway and Kochen~\cite{Aperes}, 
while Peres
provided a compact proof based on cubic
symmetry using 33
observables~\cite{Asher1991-JPhA}.
In
higher dimensions the number of observables can be
further reduced and more compact proofs are
possible~\cite{Mermin-hvt,Exp_SIC_2008}. 

Klyachko {\em et.~al.} found a minimal set of 5
observables for a qutrit for which the predicted
value for quantum correlation exceeds the bound
(the KCBS inequality) imposed by non-contextual
deterministic models~\cite{kcbs_2008}. The
violation observed is state dependent and one can
find states that do not allow for stronger than
classical correlations for the same set of
observables. A state independent violation of a
non-contextuality inequality implies that 
stronger correlations than classical are possible for all
states
for the same set
of observables~\cite{State_ind_cond_2015}. 
In a 3-dimensional Hilbert space
the minimum number of observables required to
achieve such a violation is 13~\cite{Yu_oh_2012,State_ind_cond_2015} and can be brought down to 9 if one
excludes the maximally mixed
state~\cite{9_obs_2012}. Recently graph theory has
also been used to describe contextuality
scenarios,
where vertices describe
unit vectors and 
edges describe
the orthogonality relationships
between
them~\cite{Graph_cabello_2014,Comb_acin}.

While at the level of individual measurements
quantum mechanics is contextual, the probability
distribution for an observable $A$ does not depend
upon the context and is not disturbed by other
compatible observables being measured. This is
called the `no-disturbance' principle and leads to
interesting monogamy relations for contextuality
inequalities~\cite{Context_mono_2012} similar to
those obeyed by Bell-type
inequalities~\cite{Bell_mono_2009}. These monogamy
relations are a powerful expression of quantum
constraints on correlations without involving a
tensor product structure, and we shall exploit
them in our work.

Non-trivial quantum features of the world play an
important role in quantum information
processing~\cite{Nielsen_Chuang} and in particular
in making the QKD protocols~\cite{Crypt_review}
fundamentally secure
as opposed to their classical
counterparts~\cite{Sec_qkd, Acin_sec,
Sec_monogamy}. 
QKD protocols can be categorized
into two distinct classes, namely the `prepare and
measure schemes', and the `entanglement assisted
schemes'.  In the prepare and measure schemes
whose prime example is the BB84~\cite{BB84}
protocol, one party prepares a quantum state and
transmits it to the other party who performs
suitable measurements to generate a key. On the
other hand, the entanglement assisted protocols
utilize entanglement between two parties and a
prime example of such a protocol is the Ekert
protocol~\cite{E91}. One distinct advantage in
the entanglement assisted QKD protocols is the
ability to check security based on classical
constraints on correlations between interested
parties via Bell's inequalities.  It has also been
shown that any two non-orthogonal states suffice
for constructing a QKD protocol~\cite{B92}.  The
idea has been  extended to
qutrits~\cite{3_state_crypt_Peres} to allow four
mutually unbiased bases for QKD.  Quantum
cryptography protocols have been proven to be
robust against eavesdropping and
noise~\cite{Sec_qkd, Acin_sec,
Sec_monogamy,BB84,E91,B92,3_state_crypt_Peres,
Shor_Pres_2000, Sec_BB84, Perform_BB84}.

Our focus in this work is to explore the utility
of quantum contextuality for QKD.  While
contextuality has already been exploited for
QKD~\cite{Cabello_ququart_2011}, we propose a new
QKD protocol which is based on the KCBS scenario
and the related  monogamy
relationships~\cite{kcbs_2008,Context_mono_2012}.
Our protocol falls in the class of `prepare and
measure schemes' but still allows a security check
based on  conditions on correlations shared
between the the two parties Alice and Bob. In fact
in our protocol it is the monogamy relation of
the KCBS inequality which is responsible for unconditional
security.

We first devise a QKD protocol between Alice and
Bob utilizing the KCBS scenario of contextuality
as a resource with post-processing of outcomes
allowed on Alice's site. Considering Eve as an
eavesdropper and using the novel graph theoretic approach~\cite{Graph_cabello_2014, Context_mono_2012} we then derive an appropriate monogamy
relation between Alice-Bob and Alice-Eve
correlations for the optimal settings of Eve. From
this monogamy relationship, we then explicitly
calculate the bounds on correlation to be shared
among Alice and Bob demonstrating the security of
the protocol.  Our protocol enjoys a distinct
advantage of not employing entanglement as a
resource which is quite costly to produce, and
still allows for a security test based on the
KCBS inequality which is
analogous to Bell-like test for security available
for the entanglement based protocols.
Further, it
can be transformed into an entanglement assisted
QKD protocol by making suitable adjustments.  
Although our protocol
is not device independent, 
it adds a new angle to the QKD
protocol research.

The material in the paper is arranged as follows:
In Section~\ref{back} we provide a brief review of
the KCBS inequality. In Section~\ref{the_protocol}
we describe our protocol, in
Section~\ref{monogamy} we derive the monogamy
relations for the required measurement settings
and in Section~\ref{sec} we discuss the security
of the protocol.  Section~\ref{conc} offers some
concluding remarks. 
%%%%%%%%%%%%%%%%%%%%%%%%%%%%%%%%%%%%%%%%%%%%%%%%%%%%%%
\section{KCBS inequality}
\label{back}
%%%%%%%%%%%%%%%%%%%%%%%%%%%%%%%%%%%%%%%%%%%%%%%%%%%% 
\begin{figure}[H]
\centering
\includegraphics[scale=1]{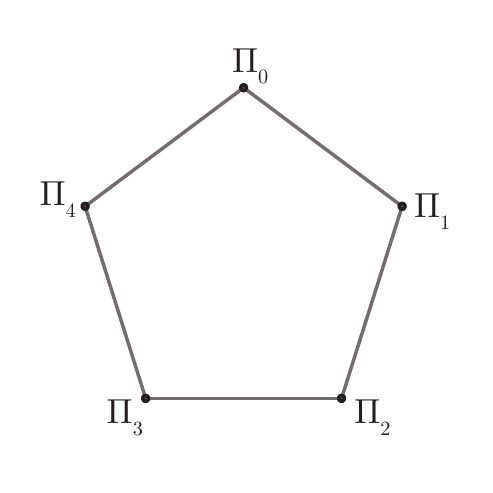}
\caption{The KCBS orthogonality graph. Each vertex
corresponds to a projector and the edge linking
two projectors indicates their orthogonality relationship.}
\label{fig:kcbsineq}
\end{figure}
%%%%%%%%%%%%%%%%%%%%%%%%%%%%%%%%%%%%%%%%%%%%%%%%%%%
The KCBS inequality is used as a test of
contextuality in systems with
Hilbert space dimension three and more. In this section we
review two equivalent formulations of the
inequality, one of which will be directly used in
our QKD protocol to be described later.

Consider a set of five observables which are
projectors in a 3-dimensional Hilbert space.  The
projectors are related via an orthogonality graph
as given in Figure~\ref{fig:kcbsineq}. The
vertices in the graph correspond to the
projectors, and two projectors are orthogonal to
each other if they are connected by an edge.  A
set of projectors which are mutually orthogonal
also commute pairwise and can therefore be
measured jointly. Such a set of co-measurable
observables is called a {\em context}. Therefore,
in the KCBS scenario, every edge between two
projectors denotes a measurement context and each
projector appears in two different contexts.
However, a non-contextual model will not
differentiate between different contexts of
a measurement and will deterministically assign
values to the vertices irrespective of the
context.

A deterministic non-contextual model must assign a
value $0$ or $1$ to the $i^{th}$ vertex and
therefore the probability that the vertex is
assigned a value $1$, denoted by $P_i$, takes
values $0$ or $1$ (and the corresponding
probability for a vertex to have value $0$
is $1-P_i$). In such a non-contextual
assignment the maximum number of vertices that can
be assigned the probability $P_{i} = 1$
(constrained by the orthogonality relations), is 2
irrespective of the state. Therefore, 
\begin{equation}
\tilde{K}(A,B)=\frac{1}{5}\sum_{i = 0}^{4}P_i \leq \frac{2}{5}.
\label{eq1}
\end{equation} 
This is the KCBS inequality~\cite{kcbs_2008, Graph_cabello_2014},
which is a state-dependent
test of contextuality utilizing these projectors,
and is satisfied by
all non-contextual deterministic models. In a
quantum mechanical description, given a quantum
state and the projector $\Pi_i$ we can calculate
the probabilities $P_i$ readily and it turns out
that the sum total probability can take values
upto $\frac{\sqrt{5}}{~5} > \frac{2}{5}$, with the maximum value
attained for a particular pure state. Therefore,
quantum mechanics does not respect non-contextual
assignments and is a contextual theory. In a more
general scenario, where one only uses the
exclusivity principle~\cite{Comb_acin} - that the sum of
probabilities for two mutually exclusive events
cannot be greater than unity - one can reach the 
algebraic maximum of the inequality namely,
\begin{equation}
\text{Max}\frac{1}{5}\sum_{i=0}^{4} P_i = \frac{1}{2}.
\end{equation} 
Unlike in inequality~(\ref{eq1}), here $P_i$s can
take continuous values in the interval  $\left[ 0,
1 \right]$.  The bounds so imposed by 
non-contextuality, quantum theory and the exclusivity principle
can be identified with graph theoretic invariants
of the exclusivity graph of the five projectors,
which in this case is also a
pentagon~\cite{Graph_cabello_2014}.

The correlation can be further analyzed if one
considers observables 
which take values $X_{i}\in\lbrace -1, +1\rbrace$ 
and are related to the projectors considered above as
\begin{equation} 
X_{i} = 2\Pi_{i} - I.
\label{eq:eq2} 
\end{equation} 
One can then reformulate Eqn.~(\ref{eq1}) in terms
of anti-correlation between two measurements
as~\cite{LSW_2011}
\begin{equation}
K(A,B)=\frac{1}{5}\sum_{i = 0}^{4}P(X_{i} \neq X_{i+1}) 
\leq \frac{3}{5}.
\label{eq:kcbs}
\end{equation}
Where $i+1$ is sum modulo 5 and $P(X_{i}\neq
X_{i+1})$ denotes the probability that a joint
measurement of $X_{i}$ and $X_{i+1}$ yields
anti-correlated outcomes. Eqn.~(\ref{eq:kcbs}) is
obeyed by all non-contextual and deterministic
models. However, quantum theory can exhibit 
violation of the above inequality. The
maximum value that can be achieved in quantum
theory is for a pure state and turns out to be 
\begin{equation}
\frac{1}{5}\sum_{i = 0}^{4}P_{\rm QM}(X_{i} \neq X_{i+1}) 
= \frac{4\sqrt{5}-5}{5}>\frac{3}{5}.
\label{eq:kcbs2}
\end{equation}
It should be noted that the maximum algebraic
value of the expression on the left hand side of
the KCBS inequality as formulated in
Eqn.~(\ref{eq:kcbs}) is one. We shall use this
formulation of the KCBS inequality directly in our
protocol in the next section as it allows
evaluation of (anti-)correlation between two
joint measurements.
%%%%%%%%%%%%%%%%%%%%%%%%%%%%%%%%%%%%%%%%%%%%%%%%%
\section{The QKD protocol, contextuality monogamy
and security}
\label{res}
\subsection{The protocol}
\label{the_protocol}
\begin{figure}[H]
\centering
\includegraphics[scale=1]{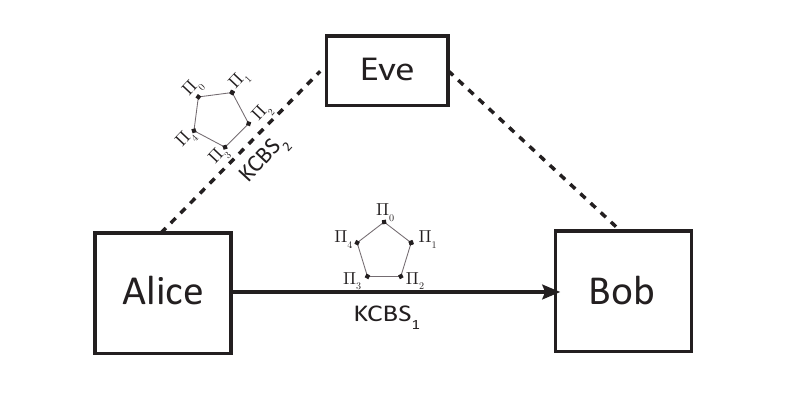}
\caption{Alice and Bob are trying to violate the
KCBS inequality [$K(A, B)$], while Eve in her
attempts to gain information is trying to violate
the same inequality with Alice [$K(A, E)$].}
\label{protocol_channel}
\end{figure}

In a typical key-distribution situation, two
separated parties Alice and Bob want to share a
secret key securely.  They both have access to the
KCBS scenario of five projectors. Alice randomly
selects a vertex $i$ and prepares the
corresponding pure state $\Pi_i$ and transmits the
state to Bob. Bob on his part, also randomly
selects a vertex $j$ and performs a measurement
$\left\lbrace \Pi_j, I-
\Pi_j\right\rbrace$ on the state. We denote $i$
and $j$ as the settings of Alice and Bob
respectively. The outcome of Bob's measurement
depends on whether he ended up measuring in the
context of Alice's state or not. The outcome
$\Pi_j$ is assigned the value $1$ and the outcome $I
- \Pi_j$ is assigned the value $0$. After the
measurement, Bob publicly announces his
measurement setting, namely the vertex $j$.  Three
distinct cases arise:
\begin{itemize}
\item[\bf C1:] $i,j$ are equal($i=j$): 
By definition Bob is assured to get
the outcome $1$. Alice notes down a $0$ with
herself and publicly announces that the
transmission was successful. Both of them
thus share an anti-correlated bit.
%%%%%%%%%%%%%%%%%%%%%%%%%%%%%%%%%%%%%%%%%%%%%
\item[\bf C2:] $i,j$ are in context but not equal: 
Bob's projector is
in the context of Alice's state. Since the state Alice
is sending is orthogonal to Bob's chosen
projector, he is assured to get the outcome $0$.
Alice then notes down $1$ with herself and
publicly announces that the transmission was
successful and Bob uses his outcome as part of the
key. This way they both share an anti-correlated
bit. It should be noted that Alice does not note
down her part of the key until Bob has announced
his choice of setting.  
%%%%%%%%%%%%%%%%%%%%%%%%%%%%%%%%%%%%%%%%%%%%%%%%
\item[\bf C3:] $i,j$ are not in context: Bob's
projector does not lie in the context of Alice's
state.  Alice publicly announces that the
transmission was unsuccessful and they try again.
However they keep this data, as it may turn out to
be useful to detect Eve. 
\end{itemize}
%%%%%%%%%%%%%%%%%%%%%%%%%%%%%%%%%%%%%%%%%%%%%
Using the protocol, Alice and Bob can securely
share a random binary key. Their success depends
on chances that Bob's measurements are made in
the context of Alice's state. Whenever Bob
measures in the correct context which happens
three-fifths of the time,
Alice is able to ensure that they
have an anti-correlated key bit. When Bob
measures in the same context but not the same
projector as Alice, she notes down a 1 with
herself and thus they share a 1-0
anti-correlation. On the other hand, when Bob
measures the same projector as Alice's state, she
notes down a 0 with herself and again they share a
0-1 anti-correlation. At no stage Alice needs to
reveal her state in public or to Bob. The QKD
scenario is depicted in Figure~\ref{protocol_channel}.

In the ideal scenario without any eavesdropper,
Alice and Bob will always get an anti-correlated
pair of outcomes and therefore will violate the
KCBS inequality to its algebraic maximum value
which is one. It should be noted that they are
able to achieve the algebraic bound because when
Bob ends up measuring the same projector as Alice, 
she notes down $0$ on her side which
is not the quantum outcome of her state. Thus
this in no way is a demonstration that quantum
theory reaches the algebraic bound of KCBS
inequality which in fact it does not. However, in
the presence of an eavesdropper the violation of
the KCBS inequality can be used as a test for
security as will be shown later. The
presence of Eve is bound to decrease the Alice Bob
anti-correlation and that can be checked by
sacrificing part of the key.

The key as generated by the above protocol
although completely anticorrelated is not
completely random, there are more ones in the key
than zeros. Therefore, the actual length of
the effective key is
smaller than the number of successful
transmissions. In order to calculate the actual
key rate we compute the Shannon information of the
transmitted string.  Given the fact that $P_0 =
\frac{1}{3}$ and $P_1 = \frac{2}{3}$ for the
string generated out of successful transmission,
the Shannon
information turns out be 
\begin{equation} S =
-P_0\log_2P_0 -P_1\log_2P_1=0.9183 
\end{equation} 
The probability of success ({\em i.e.} when Bob
chooses his measurement in the context of Alice's
state) is $\frac{3}{5}$ as stated earlier. Thus the average key
generation rate per transmission can be obtained
as $\frac{3}{5} S = 0.55 $.  
We tabulate the average key rate of a few
QKD protocols in the absence of an eavesdropper in
Table~\ref{table: comparison}.

\begin{table}[h]
\centering
\begin{tabular}{|p{2.6cm}|p{2.7cm}|p{2.7cm}|}
\hline
~QKD protocol & Success probability & 
Av. key rate in bits\\
& (per transmission) &(per transmission) \\
\hline
\hline
~BB84 (2 basis) &
~~~~~~~~~$1/2$ & $~~~~~~~~0.50$ \\
\hline
~BB84 (3 basis) & ~~~~~~~~~$1/3$ & $~~~~~~~~0.50$ \\
\hline
~Ekert(EPR pairs) & ~~~~~~~~~$1/2$ & $~~~~~~~~0.50$ \\
\hline
~3-State~\cite{3_state_crypt_Peres} &
 ~~~~~~~~~$1/4$ & $~~~~~~~~0.50$ \\
\hline
~\textbf{KCBS}  & ~~~~~~~~~$3/5$ &   $~~~~~~~~0.55$ \\
\hline
\end{tabular}
\caption{The key rate for various QKD protocols in
the absence of an eavesdropper. As can be seen
the KCBS protocol offers a little higher key rate
compared to the other protocols.} 
\label{table: comparison}
\end{table}
%%%%%%%%%%%%%%%%%%%%%%%%%%%%%%%%%%%%%%%%%%%%%%%

It is instructive to note that the above QKD
protocol can be transformed into an `entanglement
assisted' protocol, where Alice and Bob share an
isotropic two-qutrit maximally entangled state: 
\begin{equation}
|\psi\rangle = \frac{1}{\sqrt{3}}\sum_{k=0}^{2}|kk\rangle.
\end{equation}
Alice randomly chooses a measurement setting $i$
and implements the measurement
$\lbrace\Pi_i,I-\Pi_i\rbrace$ on her part of the
entangled state. In the situation when she gets a
positive answer and her states collapses to
$\Pi_i$ Bob's state collapses to $\Pi_i$ too.
This then becomes equivalent to the situation where
Alice prepares the state $\Pi_i$ and sends it to
Bob. The probability of this occurrence is $1/3$.
Bob too randomly chooses a measurement setting $j$
and implements the corresponding measurement. 
The rest of the protocol proceeds exactly as in
the case of prepare and measure scenario.

Although there are a number of possible choices
for the projectors $\Pi_i$, we detail below a
particular choice of vectors $|v_i\rangle$
(un-normalized)  
corresponding
to the projectors $\Pi_i$,
on which the above
assertions can be easily verified.
\begin{eqnarray}
|v_0\rangle &=& \left( 1, 0,
\sqrt{\cos\frac{\pi}{5}}\right) \nonumber \\
|v_1\rangle &=& \left( \cos\frac{4\pi}{5},
-\sin\frac{4\pi}{5},
\sqrt{\cos\frac{\pi}{5}}\right) \nonumber \\
|v_2\rangle &=& \left( \cos\frac{2\pi}{5},
\sin\frac{2\pi}{5},
\sqrt{\cos\frac{\pi}{5}}\right) \nonumber \\
|v_3\rangle &=& \left( \cos\frac{2\pi}{5},
-\sin\frac{2\pi}{5},
\sqrt{\cos\frac{\pi}{5}}\right) \nonumber \\
|v_4\rangle &=& \left( \cos\frac{4\pi}{5},
\sin\frac{4\pi}{5},
\sqrt{\cos\frac{\pi}{5}}\right)
\end{eqnarray}
With 
\begin{equation}
\Pi_i=\frac{\vert v_i\rangle \langle v_i
\vert}{\langle v_i \vert v_i \rangle},\quad
i=0,1,2,3,4.
\end{equation}
Thus our `prepare and measure' protocol can be
translated into an `entanglement assisted'
protocol. 
We have provided this mapping for the sake of
completeness and in 
our further discussions we will continue to
consider
the prepare and measure scheme. 
%%%%%%%%%%%%%%%%%%%%%%%%%%%%%%%%%%%%%%%%%%%%%%%%
\subsection{Contextuality monogamy}
\label{monogamy}
In quantum mechanics, given observables  $A, B,
C$, such that $A$ can be jointly measured both
with $B$ and $C$ ({\em i.e.} it is compatible with
both) the marginal probability distribution $P(A)$
for $A$ as calculated from both the joint
probability distributions $P(A,B)$ and $P(A,C)$
is the same: 
\begin{equation} 
\sum_{b} P(A=a, B=b) = \sum_c P(A=a, C=c) = P(A = a). 
\end{equation} 
This is called the `no-disturbance' principle and
it reduces to the `no-signaling' principle when
the measurements $B$ and $C$ are performed on
spatially separate systems.

The `no-disturbance' principle can be used to
construct contextuality monogamy relationships of a
set of observables if they can be partitioned into
disjoint subsets each of which can reveal
contextuality by themselves but cannot be
simultaneously used as tests of contextuality.
%%%%%%%%%%%%%%%%%%%%%%%%%%%%%%%%%%%%%%%%%%%%%%%%%%
\begin{figure}[H]
\centering
\includegraphics[scale=1]{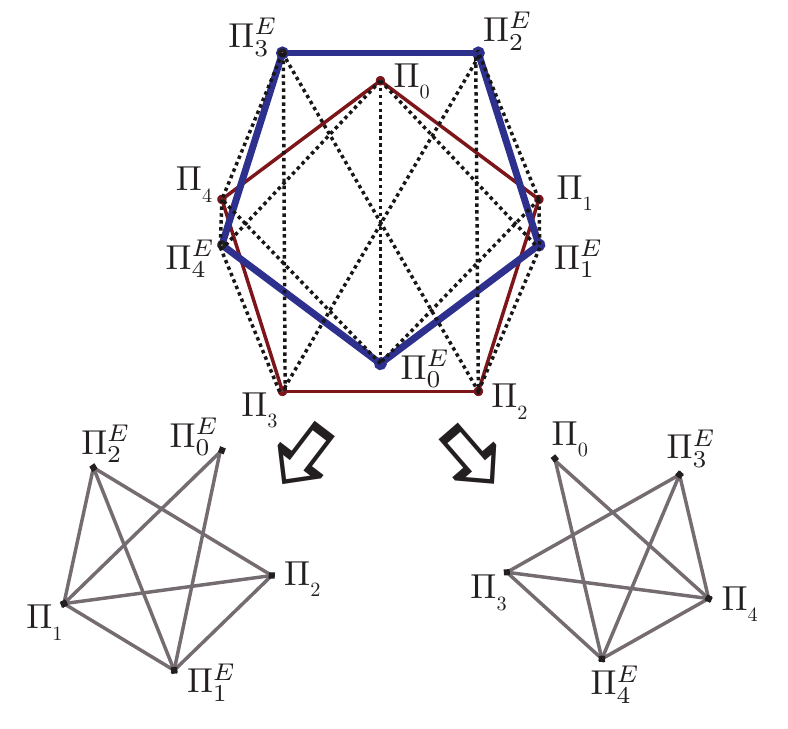}
\caption{Joint commutation graph (top) of
Alice-Bob KCBS test (Thin-red) and Alice-Eve KCBS
test (Thick-blue) and its decomposition into two
chordal subgraphs (below). Dotted edges indicate
commutation relation between two projectors
belonging to the two different KCBS tests. (color
online)} \label{fig:mono}
\end{figure}

Consider the situation where Alice and Bob are
different parties who make preparations and
measurements as detailed in
Section~\ref{the_protocol}. We consider the
possibility of a third party Eve who tries to
eavesdrop on the conversation between them. As
will be detailed in Section~\ref{sec} Eve will
have to violate the KCBS inequality with Alice to
gain substantial  information about the key.

We denote the Alice-Bob KCBS test by
$\tilde{K}(A,B)$ with projectors
$\left\lbrace \Pi_i \right\rbrace$ and Alice-Eve
KCBS test by $\tilde{K}(A,E)$ with
projectors $\left\lbrace \Pi^E_i \right\rbrace$.
We have assumed different projectors in the
two KCBS tests for clarity in derivation of a
monogamy relationship, but essentially the
measurements to be performed by Eve would have to
be the same as that of Bob to mimic 
Alice and Bob's KCBS scenario as
will be detailed in Section~\ref{sec} where we
take up  the security analysis  of our protocol. In this
joint scenario the $\Pi_i^{th}$ projector is
connected by an edge to $\Pi_{i+1}$,
$\Pi^E_{i+1}$, $\Pi_{i-1}$, $\Pi^E_{i-1}$ and
$\Pi^E_i$, where $i+1$ and $i-1$ are taken modulo
5 and the presence of an edge denotes
commutativity between the two connected vertices.
These relationships follow from the fact that the
projectors used by Eve will follow the same
commutativity relationships as the original KCBS
scenario. By introducing herself in the channel,
Eve has created an extended scenario which will
have to obey contextuality monogamy due to the
no-disturbance principle.  The no-disturbance
principle guarantees that the marginal
probabilities as calculated from the joint
probability distribution do not depend on the
choice of the joint probability distribution used.

We follow the graph theoretical approach
developed to derive generalized monogamy
relationships based only on no-disturbance
principle in reference~\cite{Context_mono_2012}. A joint
commutation graph representing a set of $n$
KCBS-type inequalities each of which has a
non-contextual bound $\alpha$ gives rise to a
monogamy relationship if and only if its vertex
clique cover number is $n.\alpha$. The vertex
clique cover number is the minimum number of
cliques required to cover all the vertices of the
graph and a clique is a graph in which all
non-adjacent vertices are connected by an edge.
The joint commutation graph considered in the
protocol resulting in the presence of Eve
satisfies the condition for the existence of a
monogamy relationship between Alice-Bob and
Alice-Eve KCBS inequalities as can be seen from
Fig.~\ref{fig:mono}.

In order to derive the monogamy relationship one
needs to identify $m$ chordal sub-graphs of the
joint commutation graph such that the sum of their
non-contextual bounds is $n.\alpha$. A chordal
graph is a graph which does not contain induced
cycles of length greater than 3. As shown
in reference~\cite{Context_mono_2012} a chordal graph admits
a joint probability distribution and therefore
cannot violate a contextuality inequality. To this
end we identify the decomposition of the joint
commutation graph into two chordal subgraphs such
that each vertex appears at most once in both the
subgraphs, as shown in Fig.~\ref{fig:mono}. Their
maximum non-contextual bound will then be given by the
independence number of the subgraph. Therefore,
\begin{eqnarray}
p(\Pi^E_0) + p(\Pi_2) + p(\Pi^E_1) + p(\Pi_1) + 
p(\Pi^E_2) \leq 2, \\
p(\Pi_0) + p(\Pi_3) + p(\Pi^E_3) + p(\Pi_4) + 
p(\Pi^E_4)  \leq 2. 
\end{eqnarray} 
Adding and grouping the terms according to their
respective inequalities
(Eqn.(\ref{eq1})) and normalizing, we get
\begin{equation}
\tilde{K}(A, B) + 
\tilde{K}(A, E)\leq \frac{4}{5}.
\label{eq:mono1}
\end{equation}
If the projectors involved in the KCBS tests are
transformed according to Eqn.(\ref{eq:eq2}), then
using the KCBS given in Eqn.(\ref{eq:kcbs}) 
the monogamy relationship reads as 
\begin{equation}
K(A, B) + K(A, E) \leq \frac{6}{5}.
\label{eq:mono3}
\end{equation}
The relationship derived above follows
directly from the no-disturbance principle and
cannot be violated. In other words, the
correlation between  Alice and Eve is
complementary to the correlation between Alice
and Bob and thus if one is strong the other has to
be weak.  One can thus use this fundamental
monogamous relationship to derive conditions for
unconditional security as will be shown in the
next section.    
%%%%%%%%%%%%%%%%%%%%%%%%%%%%%%%%%%%%%%%%%%%%%%%%%
\subsection{Security analysis}
\label{sec}
In this section we prove that the above QKD
protocol is secure against
individual attacks by an eavesdropper Eve. We
first motivate the best strategy available to an
eavesdropper limited only by the no-disturbance
principle. The best strategy would then dictate
the optimal settings to be used to maximize the
information of Eve about the key. We then prove
unconditional security of the protocol based on
monogamy of the KCBS inequality. The analysis
is inspired by the security proof for QKD
protocols based on the monogamy of violations of
Bell's inequality~\cite{Sec_monogamy}.

Alice and Bob perform the protocol a large
number of times and share the probability
distribution $P(a, b|i, j)$, which denotes the
probability of Alice and Bob obtaining outcomes $a,
b \in \lbrace 0, 1\rbrace$ when their settings are
$i, j \in \lbrace 0, 1, 2, 3, 4 \rbrace$
respectively. In the ideal case they obtain $a
\neq b$ when $j = i + 1$, where addition is taken
modulo 5. 
However in the presence of Eve, the secrecy of
correlation between Alice and Bob has to be
ensured even if Eve is distributing the
correlation between them. On the other hand, Eve
would like to obtain information about the
correlation between Alice and Bob and the
associated key.
Eve
can attempt to accomplish this in several ways which might
include intercepting the information from Alice
and re-sending to Bob after gaining suitable
knowledge about the key. It could also be that she
is correlated to Alice's preparation system or to
Bob's measurement devices. In other words Eve has
access to a tripartite probability distribution
$P(a, b, e|i, j, k)$, where Alice, Bob and Eve
obtain outcomes $a$, $b$ and $e$ when their
settings are $i$, $j$ and $k$ respectively. It is
required that the marginals to this probability
distribution correspond to the observed
correlation between Alice and Bob as will be
shown below. In general it is not easy to
characterize the strategy of an eavesdropper
without placing some constraints on her.

For the following security analysis we place
fairly minimal restrictions on the eavesdropper. It is
required of her to obey the no-disturbance
principle and as a consequence her correlation
with Alice will be limited by
monogamy~(\ref{eq:mono3}). Such a constraint is
well motivated because it is a fundamental law of
nature and will have to be obeyed at all times.

We assume that the correlation observed by
Alice and Bob, $P(a, b|i, j)$ as defined above,
is a consequence of marginalizing over an extended
tripartite probability distribution $P(a, b, e|i,
j, k)$, distributed by an eavesdropper Eve:
\begin{equation} \begin{aligned} P(a, b|i, j) &=
\sum_e P(a, b, e|i, j, k) \\ &=\sum_e P(e|k)P(a,
b|i, j, k, e).  \end{aligned} 
\label{eq:abcorrel}
\end{equation} Where the second equality is a
consequence of the no-disturbance principle: Eve's
output is independent of the settings used by
Alice and Bob. We can also analyze the
correlation between Alice and Eve in a similar
manner: \begin{equation} \begin{aligned} P(a, e|i,
k) &= \sum_b P(a, b, e|i, j, k) \\ &=\sum_b
P(b|j)P(a, e|i, j, k).  \end{aligned} 
\label{eq:aecorrel} 
\end{equation} 
Where the second equality
also follows from the no-disturbance principle and
implies that Eve can decide on her output based on
the settings disclosed by Bob. Bob's outcome,
however, cannot be used as it is never disclosed
in the protocol. The natural question that arises
now is how strong does the correlation between
Alice and Bob need to be such that the protocol is
deemed secure. As will be seen the question can be
answered by monogamy of contextuality.

The QKD scenario now is as follows: Alice and
Bob utilize the preparations and measurements as
detailed in Section~\ref{the_protocol}, while an
eavesdropper Eve limited only by the
no-disturbance principle is assumed to distribute
the correlation between them. Whenever Eve
distributes the correlation between herself and
Alice she uses the same measurement settings as
Bob to guess the bit of Alice. This way Eve can
hope to gain some information about the key.
However, contextuality monogamy limits the amount
to which Eve
can be correlated to Alice without
disturbing the correlation between Alice and Bob
significantly as shown in Section~\ref{monogamy}.

The condition for a secure key distribution
between
Alice and Bob in terms of Alice-Bob  mutual
information 
$I(A : B)$  
and Alice-Eve mutual information 
$I(A : E)$ is~\cite{secure_cond}:
\begin{equation}
I(A : B) > I(A : E).
\end{equation}
For individual attacks and binary outputs
of Alice 
it essentially means that
the probability
$P_\text{B}$ that Bob guesses the bit of Alice
should be greater than the probability,
$P_\text{E}$ for Eve to correctly guess the bit of
Alice. 
Thus the above condition simplifies to~\cite{sec_bell_reply}
\begin{equation}
P_\text{B} > P_\text{E}.
\label{sec_cond}
\end{equation} 
Bob can correctly guess the bit of Alice
with probability $P_\text{B} = K(A, B)$. For
$K(A, B) = 1$ Bob has perfect knowledge
about the bit of Alice while for $K(A, B) =
0$ he has no knowledge. For any other values of
$K(A, B)$ they may have to perform a security
check.

We assume that Eve has a
procedure that enables her to distribute
correlation according to
Eqns.~(\ref{eq:abcorrel}-\ref{eq:aecorrel}).
The procedure takes an input $k$ among
the five possible inputs according to the KCBS
scenario and outputs $e$.
She uses this outcome to
determine the bit of Alice when Alice's setting was
$i$. The probability that Eve correctly guesses
the bit of Alice is denoted by $P_{ik}$. Since
there are 5 possible settings for Alice and Eve
each, the
average probability for Eve to be successful
$P_\text{E}$ is,
\begin{eqnarray}
P_\text{E} &=& \frac{1}{15}\sum_{i=0}^{4} \left(
P_{ii} + P_{ii+1}+P_{ii-1}\right)\nonumber \\
&\leq& \text{max}\lbrace P_{ii}, P_{ii+1},
P_{ii-1}| \forall\,  i \rbrace.  
\label{eve_success}
\end{eqnarray}
The terms in the above equation
denote the success probability of Eve when she
uses the same setting as Alice and when she
measures in the context of Alice, respectively. 
For
all other cases she is unsuccessful. Without loss
of generality we can assume that $P_{01}$ is the
greatest term appearing in Eqn~(\ref{eve_success}). This
corresponds to the success probability of Eve when
her setting is $1$ and Alice's setting is $0$.
However, Alice's setting is not known to Eve as it
is never disclosed in the protocol. Therefore the
best strategy that Eve can employ is to always
choose her setting to be $1$ irrespective of
Alice's settings and try to violate the KCBS
inequality with her. The probabilities that appear
in the KCBS inequality would then be,
\begin{eqnarray}
P(a \neq e|i=0, k=1) &=& P_{01}=P_{01}, 
\nonumber \\
P(a \neq e|i=1, k=1) &=& P_{11}=1 - P_{01},
\nonumber  \\
P(a \neq e|i=2, k=1) &=& P_{21}\leq P_{01}, \nonumber \\
P(a \neq e|i=3, k=1) &=& P_{31}\leq P_{01}, \nonumber \\
P(a \neq e|i=4, k=1) &=& P_{41}\leq P_{01}.
\end{eqnarray}
The probability for Eve to get a particular
outcome is independent of Alice's choice of
settings. Her best strategy to eavesdrop can at
most yield all the preceding probabilities to be
equal (except the second term which will show a
correlation instead of the required
anti-correlation) which will maximize $K(A, E)$.
Evaluating the KCBS violation for Alice and Eve,
we get,
\begin{equation}
K(A, E) = \frac{3}{5}P_{01} +
\frac{1}{5}>\frac{3}{5}P_\text{E} + \frac{1}{5}.
\end{equation}
Using the monogamy relationship given by Eqn.~(\ref{eq:mono3}), we get,
\begin{equation}
\frac{3}{5}P_\text{E} + \frac{1}{5}\leq \frac{6}{5} - P_\text{B}.
\end{equation}
For the protocol to work 
Eqn.~(\ref{sec_cond}) must hold  and the above
condition implies that it happens only if
\begin{equation}
K(A, B) > \frac{5}{8}.
\label{condition}
\end{equation}
Therefore the protocol is unconditionally secure
if Alice and Bob share KCBS correlation greater
than $\frac{5}{8}$. It is worth mentioning that
$\frac{5}{8}$ is lesser than the maximum violation
of the KCBS inequality in quantum theory.

As shown in reference~\cite{Context_mono_2012} the
monogamy relation~(\ref{eq:mono3}) is a minimal
condition and no stronger conditions exist.
This implies that any QKD protocol
whose security is based on the violation of the
KCBS inequality cannot offer security if the
condition given in Eqn.~(\ref{condition}) is not satisfied. This
quantifies the minimum correlation required for
unconditional security. We conjecture that no key
distribution scheme based on the violation of the
KCBS inequality can perform better than our
protocol since we utilize post-processing on
Alice's side to extend the maximum violation of
the KCBS inequality upto its algebraic maximum.
%%%%%%%%%%%%%%%%%%%%%%%%%%%%%%%%%%%%%%%%%%%%%%%%%
\section{Conclusions} 
\label{conc} 
The cryptography protocol we presented is a direct
application of the simplest known test of
contextuality namely the KCBS inequality and the
related monogamy relation. For the protocol
to work, Alice and Bob try to achieve the maximum
possible anti-correlation amongst themselves.
They achieve the algebraic maximum of the KCBS
inequality by allowing post-processing on Alice's
site. We then showed that any eavesdropper will
have to share a monogamous relationship with Alice
and Bob severely limiting her eavesdropping. For
this purpose we derived a monogamy relationship
for the settings of Eve which allow her to gain
optimal information. We found that the optimal
information gained by Eve cannot even allow her to
maximally violate the KCBS inequality as allowed
by quantum theory. Such an unconditional security
provides a significant advantage to our protocol
since it does not utilize the costly resource of
entanglement. Furthermore, being a prepare and
measure scheme of QKD it also allows
for a check of security via the violation of the
KCBS inequality much like the protocols based on
the violation of Bell's inequalities. 
Finally, we note that our protocol is a
consequence of contextuality monogamy
relationship, which are expected to play an
interesting role in quantum information
processing.

%%%%%%%%%%%%%%%%%%%%%%%%%%%%%%%%%%%%%%%%%%%%%%%%%%
\begin{acknowledgments} We acknowledge Sandeep K.
Goyal and Adan Cabello for useful discussions.
Arvind acknowledges funding from DST India under
Grant No. EMR/2014/000297 and Kishor Bharti
acknowledges the KVPY fellowship of DST India.
\end{acknowledgments}
%\bibliography{crypt} 
\end{document}